\documentclass[11pt]{article}
\begin{document}
\pagestyle{plain}
\huge
\title{\bf Physics of  particles  in the rotating tube}
\large
\author{Miroslav Pardy\\[7mm]
Department of Physical Electronics \\
and\\
Laboratory of Plasma physics\\[5mm]
Masaryk University \\
Kotl\'{a}\v{r}sk\'{a} 2, 611 37 Brno, Czech Republic\\
e-mail:pamir@physics.muni.cz}
\date{\today}
\maketitle
\vspace{25mm}

\begin{abstract}
The classical and the quantum motion of a massive body in the rotating tube is considered. Photon is included. The spin motion described by the  Bargmann-Michel-Telegdi equation is considered in the  rotation tube  and rotating system.
\end{abstract}

\vspace{3mm}


\vspace{13mm}

\section{Introduction}
The accelerators accelerate only charged particles. It is surprising that neutral particles such as photons, neutrons and so on can be accelerated by rotating tube. We show the mechanics and quantum mechanics of the motion of such particle in the rotating tube. We also include the general relativistic view on the rotating plane  in order to generalize Bargmann-Michel-Telegdi equation for rotating systems. In order to see the difference between physics in rotation tube and in accelerating systems, let us first consider the massive body in the noninertial frame.

The specific characteristics of the mechanical systems in the rotating framework follow from  the differential equations describing the massive body in the noninertial systems (Landau et al., 1965). We will see later that the motion of a body in the rotating tube cannot be described by the formalism for rotating disk. We start by the text of Landau et al. (1965). 

Let be the Lagrange function of a point particle in the inertial
system as follows:

$$L_{0}  = \frac{m{\bf v}_{0}^{2}}{2} - U \eqno(1)$$
with the following equation of motion

$$ m\frac{d{\bf v}_{0}}{dt} = -\frac{\partial U}{\partial {\bf r}},
\eqno(2)$$
where the quantities with index $0$ correspond to the inertial system.

The Lagrange equations in the noninertial system is of the same form as
that in the inertial one, or,

$$\frac{d}{dt}\frac{\partial L}{\partial {\bf v}} =  \frac{\partial
  L}{\partial {\bf r}}.\eqno(3)$$

However, the Lagrange function in the noninertial system is not the
same as in eq. (1) because it is transformed.

Let us first consider the system $K'$ moving relatively to the system
$K$ with the velocity ${\bf V}(t)$. If we denote the velocity of a
particle with regard to system $K'$ as ${\bf v}'$, then evidently

$${\bf v}_{0} = {\bf v}' + {\bf V}(t).\eqno(4)$$

After insertion of eq. (4) into eq. (1), we get

$$L_{0}'  = \frac{m{\bf v'}^{2}}{2} + m {\bf v'}{\bf V} +
\frac{m}{2}{\bf V }^{2} - U. \eqno(5)$$

The function ${\bf V }^{2}$ is the function of time only and it can be
expressed  as the total derivation of time of some new function. It
means that the term with the total derivation
in the Lagrange function can be removed from the Lagrangian. We also have:

$$ m {\bf v'}{\bf V}(t) = m{\bf V}\frac{d{\bf r'}}{dt} = \frac{d}{dt}
 (m {\bf r'}{\bf V}(t)) - m{\bf r'}\frac{d{\bf V}}{dt}.\eqno(6)$$

After inserting the last formula into the Lagrange function and after
removing the total time derivation we get

$$L' = \frac{mv'^2}{2} - m{\bf W}(t){\bf r'} - U, \eqno(7)$$
where ${\bf W} = d{\bf V}/dt$ is the acceleration of the system $K'$.

The Lagrange equations following from the Lagrangian (7) are as
follows:

$$ m\frac{d{\bf v}'}{dt} = -\frac{\partial U}{\partial {\bf r}'}  -
m{\bf W}(t).\eqno(8)$$

We see that after acceleration of the system $K'$ the new force $m{\bf W}(t)$
appears. This force is fictitious one because it is not
generated by the internal properties of some body.

In case that the system $K'$ rotates with the angle
velocity $\bf \Omega$ with regard to the system $K$, vectors  $\bf v$ 
and $\bf v'$ are related as (Landau et al., 1965)

$${\bf v}'  =  {\bf v} + {\bf \Omega}\times{\bf r}.\eqno(9)$$

The Lagrange function for this situation  is  (Landau et al., 1965 )

$$L = \frac{mv^2}{2} - m{\bf W}(t){\bf r} - U + m {\bf v}\cdot({\bf \Omega}
\times{\bf r}) + \frac{m}{2}({\bf \Omega}\times{\bf r})^{2}.\eqno(10)$$

The corresponding  Lagrange equations for the last Lagrange function are as follows (Landau et al., 1965 ):

$$ m\frac{d{\bf v}}{dt} = -\frac{\partial U}{\partial {\bf r}}  -
m{\bf W} + m{\bf r}\times\dot{\bf \Omega} + 2m{\bf v}\times{\bf \Omega}
+ m{\bf \Omega}\times \left({\bf r}\times{\bf \Omega}\right).\eqno(11)$$

 We observe in eq. (11) three so called inertial forces. The force
$m{\bf r}\times\dot{\bf \Omega}$ is connected with the nonuniform
rotation of the system $K'$ and the forces $2m{\bf v}\times{\bf
\Omega}$ and $m{\bf \Omega}\times {\bf r}\times{\bf \Omega}$
correspond to the uniform rotation. The force $2m{\bf v}\times{\bf\Omega}$
is so called the Coriolis force and it depends on the velocity of a
particle. The force $m{\bf \Omega}\times {\bf r}\times{\bf \Omega}$ is
called the centrifugal force. It is perpendicular to the rotation axes
and the magnitude of it is $m\varrho\omega^{2}$, where $\varrho$ is
the distance of a particle from the rotation axis.

\section{The massive point moving in the rotating tube}
Let us consider a force $\bf F$ acting at
a massive body  with mass $m$, where the mathematical form of this force is as follows:

$$ F_{i}= m C\varepsilon_{ijk}x_{j}\Omega_{k},\eqno(12)$$
where the dimensionality of this quantity is ${\rm kg. m}.{\rm s}^{-2}$ if $C, \Omega$ have the dimensionality of frequency. We can easily reduce the formula (12) to the vectorial form

$$ {\bf F} =  m \Omega ({\bf r}\times {\bf \Omega}),\eqno(13)$$ 
and ${\bf \Omega}$ is supposed to be angular velocity and ${\bf r}$ is the radius vector of  the position of the body with mass $m$. The force defined in such a way is perpendicular to the angular velocity ${\bf \Omega}$ and to the radius vector ${\bf r}$ and it can be physically interpreted as force acting by rotating tube on the massive body which motion is restricted to the motion inside of the tube AB, where the tube AB rotates in the x-y plane in such a way that point A is $A\equiv 0(0,0,0)$. The force (13) is formally similar to the Lorentz force acting on a charged particle moving in the constant magnetic field, but the physical meaning is diametrically different from the Lorentz force.

The equation of motion under the force (13) is evidently as follows:

$$m{\ddot {\bf r}} = m{\Omega}({\bf r}\times {\bf \Omega}) = m{\Omega}
\left |
\begin{array}{ccc}
{\bf i}& {\bf j}& {\bf k}\\
x& y& z\\
0& 0& {\Omega}  
\end{array}
\right |.
\eqno(14)$$ 

If we write for the circular motion in the x-y plane 

$${\bf r} = r({\bf i}\cos\Omega t + {\bf j}\sin\Omega t + {\bf k}z), \eqno(15)$$ 
 we get the equations of motion in the form
 
 $$\ddot x = -r\Omega^{2}\cos\Omega t, \quad \ddot y = r\Omega ^{2}\sin\Omega t, \eqno(16)$$ 
from which follows the differential equations for $r$:

$$\ddot r = \Omega^{2}r \eqno(17)$$
with the solution 

$$r = c_{1}e^{\Omega t} + c_{2}e^{-\Omega t}. \eqno(18)$$

Supposing $r(0)= r_{0}, \dot r(0) = 0$, we get the solution in the form

$$r(t) = r_{0}\cosh\Omega t. \eqno(19)$$

In case that $r(0)= r_{0}, \dot r(0) = v$, we get the solution in the form

$$r(t) = r_{0}\cosh\Omega t + \frac{v}{\Omega}\sinh\Omega t. \eqno(20)$$

Equations (19), or (20) can be immediately applied to the situation, where the tube is joined with the Earth (North Pole), where the frequency of rotation is $\Omega = 1/{\rm day} $. If we put $r_{0}= 1 {\rm m}, t = {\rm day}$, then with regard to the formula $\cosh x = 1 + x^{2}/2! + x^{4}/4! + ...$, we get $r \approx 1,5 {\rm m}$. So, we see that the rotation of Earth can be confirmed by the experiment with the rotating tube. The experiment, if performed, is the physical proof of the Earth rotation.
This experiment was not considered in the textbooks on mechanics including the Euler famous opus "Teoria motus corporum solidorum seu rigidorum". (Euler, 1790). Only Foucault pendulum is discussed (Pardy, 2007).
 
Let us remark that we can identify the equation $\ddot r = \Omega^{2}r$ with the equation for harmonic oscillator if we put $\Omega \rightarrow i\Omega$. At the same case we can say that this equation follows immediately from the incorrect physical assumption that the motion of a point particle, in a rotating tube is caused by the centrifugal force $F = m\Omega^{2}r$ to which corresponds the "potential energy" 

$$W = \int_{r_{0}}^{r}Fdr = \frac{1}{2}m\Omega^{2}(r^{2} - r_{0}^{2}).\eqno(21)$$

The kinetic energy at point $r$, at the direction of a tube, is (as follows from eq. (19))

$$E_{kin} = \frac{1}{2}mv^{2} = \frac{1}{2}m\Omega^{2}(r^{2} - r_{0}^{2}).\eqno(22)$$

So, $W = E_{kin}$. However, It is evident that there is no real centrifugal force inside of the rotating tube.  

The rotating tube in the form of the carbon nanotube can be applied to intercalate different atoms and molecules. The carbon nanotube with such intercalated atoms and molecules has new nonexpected physical properties including superconductivity behavior. So, the substantial ingredient of every science, surprise, is established. 

\section{The  motion of a photon in a  rotating tube} 

The mass of moving photon is not zero
bat is given by the Einstein relation $m_{\gamma}  = (\hbar\omega)/c^{2},$
where $\omega$ is the frequency of he photon an $c$ is the velocity of photon in vacuum. We consider the tube rotating in vacuum and the
initial of photon velocity is  $c$. With regard to the fact that the
photon velocity is constant in the rotating tube, the result of the rotation is the change of frequency of photon. Or, the final frequency is 

$$\hbar\omega' = \hbar \omega + \Delta E_{\gamma},
 \eqno(23)$$ 
where $E_{\gamma}$ is the additional energy of photon which is obtained by photon under the process of acceleration along the the trajectory of photon in the tube AB. Using equation (21), the equation for the shift of photon frequency (23) can be expressed as 

$$\hbar\omega' = \hbar \omega + \Delta E_{\gamma} = \hbar\omega + \frac{1}{2}\left(\frac{\hbar\omega}{c^{2}}\right)\Omega^{2}(r^{2}-r_{0}^{2}), \eqno(24)$$ 
 
 Although such derivation
of the change of the photon frequency is heuristical, it is necessary because there is still no theory of photons in the rotating tube. 

In case that we consider the situation, where photon is moving from B to A, then we get red shift of the frequency, which of course cannot be considered as the analogy of the red shift of the rotating meta-galaxy.

The second possibility of derivation of the change of photon frequency in the rotating tube is to consider the tube as wave guide and then to calculate the electromagnetic field in the rotating wave guide. 

\section{Motion of the spin-vector in a rotating tube}

We suppose here that it is possible to use the problem of motion of the spin in a rotating tube as an analogy with the problem of the spin-vector motion in classical relativistic mechanics presented by Bargmann, Michel and Telegdi (Berestetzkii et al., 1989). They derived so called BMT equation for motion of spin in the electromagnetic field, in the form 

$$\frac{da_{\mu}}{ds} = \alpha F_{\mu\nu}a^{\nu} + \beta v_{\mu}F^{\nu\lambda}v_{\nu}a_{\lambda},\eqno(25)$$
where $a_{\mu}$ is so called axial vector describing the classical
spin, $v_{\mu}$  is velocity and  constants $\alpha$ and $\beta$ were determined after the comparison of the postulated equations with the 
non-relativistic quantum mechanical limit. The result of such comparison is the final form of so called BMT equations:

$$\frac{da_{\mu}}{ds} = 2\mu F_{\mu\nu}a^{\nu} -2\mu'v_{\mu}F^{\nu\lambda}v_{\nu}a_{\lambda},\eqno(26)$$
where $\mu$ is magnetic moment of electron following directly from the Dirac equation and $\mu'$ is anomalous magnetic moment of electron which can be calculated as the radiative correction to the interaction of electron with electromagnetic field and follows from quantum electrodynamics.

 The BMT equation has more earlier origin. The first attempt to describe the spin motion in electromagnetic field using the special theory of relativity was performed by Thomas (1926). However, the basic ideas on the spin motion was established by Frenkel (1926, 1958). After appearing the Frenkel basic article,
many authors published the articles concerning the spin motion (Ternov et al., 1980; Tomonaga, 1997). At present time, spin of electron is its  physical attribute which follows only from the Dirac equation.

It was shown by Rafanelli and Schiller (1964), (Pardy, 1973) 
that the BMT equation can be derived from the classical limit, i.e. from the WKB solution of the Dirac equation with the anomalous magnetic moment.

If we introduce the average value of the vector of spin in the rest system by the quantity $\mbox {\boldmath $\zeta$}$,
 then the 4-pseudovector  $a^{\mu}$  is of the from 
$a^{\mu} = (0, \mbox {\boldmath $\zeta$})$ (Berestetzkii et al., 1989; Pardy, 2009). The momentum four-vector of a particle is $p ^{\mu} = (m, 0)$ in the rest system of a particle. Then the equation $a^{\mu}p_{\mu} = 0 $
is valid not only in the rest system of a particle but in the arbitrary system as a consequence of the relativistic invariance. The following general formula is also valid in the arbitrary  system
$a^{\mu}a_{\mu} = - \mbox {\boldmath $\zeta$}^{2}$.

The components of the axial 4-vector $a^{\mu}$ in the reference system, where particle is moving with the velocity ${\bf v} = {\bf p}/\varepsilon$ can be obtained by application of the Lorentz transformation to the rest system and they are as follows (Berestetzkii et al., 1989):

$$a^{0} = \frac{|{\bf p}|}{m}\mbox {\boldmath $\zeta$}_{\parallel}, \quad {\bf a}_{\perp} = \mbox {\boldmath $\zeta$}_{\perp}, \quad a_{\parallel} = \frac{\varepsilon}{m}\mbox {\boldmath $\zeta$}_{\parallel}, \eqno(27)$$
where suffices $\parallel, \perp$ denote the components of ${\bf a}$, $\mbox {\boldmath $\zeta$}$ parallel and perpendicular to the direction ${\bf p}$. The formulas for the spin components can be also rewritten in the more compact form as follows (Berestetzkii et al., 1989):

$${\bf a} = {\mbox {\boldmath $\zeta$}} + \frac{{\bf p}({\mbox {\boldmath $\zeta$}}{\bf p})}{m(\varepsilon + m)}, \quad a^{0} = \frac{{\bf a}{\bf p}}{\varepsilon} = \frac{{\mbox {\boldmath $\zeta$}}{\bf p}}{m}, \quad
{\bf a}^{2} = {\mbox {\boldmath $\zeta$}}^{2} + \frac{({\bf p}{\mbox {\boldmath $\zeta$}})^{2}}{m^{2}}.\eqno(28)$$

The equation for the change of polarization can be obtained immediately from the BMT equation in the following form (Berestetzkii et al., 1989):

$$ \frac{d{\bf a}}{dt} = \frac{2\mu m}{\varepsilon}{\bf a}\times{\bf H} + \frac{2\mu m}{\varepsilon}({\bf a}{\bf v}){\bf E} - \frac{2\mu' \varepsilon}{m}{\bf v}({\bf a}{\bf E})\; + $$

$$+ \frac{2\mu'\varepsilon}{m}{\bf v}({\bf v}({\bf a}\times {\bf H})) +  \frac{2\mu'\varepsilon}{m}{\bf v}({\bf a}{\bf v})({\bf v}{\bf E}), \eqno(29)$$
where we used the relativistic relations $c =1$, $ds = dt\sqrt{1 - v^{2}}$ , $\varepsilon = m\sqrt{1 - v^{2}}$ and the following components of the electromagnetic field (Landau et al., 1988):

$$F^{\mu\nu} = \left(\begin{array}{cccc}
0 & -E_{x} & -E_{y} & -E_{z}\\
E_{x} & 0 & -H_{z} & H_{y}\\
E_{y} & H_{z} & 0 & -H_{x}\\
E_{z} & -H_{y} & H_{x} & 0\\
\end{array} \right) \stackrel {d}{=} ({\bf E}, {\bf H}); \quad F_{\mu\nu} = ({-\bf E }, {\bf H}).\eqno(30)$$

Inserting equation ${\bf a}$ from eq. (28) into  eq. (29) and using equations 

$${\bf p} = \varepsilon{\bf v}, \quad \varepsilon^{2} = {\bf p}^{2} + m^{2}, \quad \frac{d{\bf p}}{dt} = e{\bf E} + e({\bf v}\times {\bf H}),\quad
\frac{d\varepsilon}{dt} = e({\bf v}{\bf E}),\eqno(31)$$
we get after long but simple mathematical operations the following equation for the polarization $\mbox {\boldmath $\zeta$}$

$$\frac{d {\mbox {\boldmath $\zeta$}}}{dt} = 
\frac{2\mu m  + 2\mu'(\varepsilon - m)}{\varepsilon}{\mbox {\boldmath $\zeta$}}\times {\bf H} \quad + $$

$$\frac{2\mu' \varepsilon}{\varepsilon + m}({\bf v}{\bf H})({{\bf v}\times \mbox {\boldmath $\zeta$}}) + \frac{2\mu m  + 2\mu' \varepsilon}{\varepsilon + m}{\mbox {\boldmath $\zeta$}}\times ({\bf E}\times {\bf v}).\eqno(32)$$

The equation of motion of spin in electric field as far as first
order terms in velocity $v$ is obtained from eq. (32) in the form 

$$\frac{d {\mbox {\boldmath $\zeta$}}}{dt} =
(\mu  + \mu'){\mbox {\boldmath $\zeta$}}\times ({\bf E}\times {\bf  v}) = 
\left(\frac{e}{2m} + 2\mu'\right)
{\mbox {\boldmath $\zeta$}}\times ({\bf E}\times {\bf  v}).\eqno(33)$$

It follows from equation (31) that $e\bf E$ is the electric force
interacting with spin of an electron. If we want to express eq. (14) as
the equation of spin motion in the rotating tube, then it is easy to
show that $\bf E$ must be identified by ${\bf F}/e$. Or, 

$$\frac{d {\mbox {\boldmath $\zeta$}}}{dt} =
\frac{1}{e}(\mu m  + \mu'){\mbox {\boldmath $\zeta$}}\times ({\bf F}\times {\bf  v}) = 
\frac{1}{e}\left(\frac{e}{2m} + 2\mu'\right)
{\mbox {\boldmath $\zeta$}}\times ({\bf F}\times {\bf  v}).\eqno(34)$$

The equation was never derived in the framework of the general theory of relativity and gravitation. The force which causes motion of spin is in case of the rotation tube the electric force and not the gravitational force. There is not principle equivalence between electric field and gravity.
 
\section{Quantum mechanics of a particle in a rotating tube}

The rigorous formulation of the problem of quantum mechanical motion of a charged particle in a rotating tube is to consider the situation of a charged particle  where the motion is restricted by the moving boundary conditions. Such problem was still not defined in quantum mechanical monographs or solved, or published. So we here use the heuristic approach which represents the most simple approach to the problem.

The elementary solution is to consider quantity $V(r) = \frac{1}{2}\Omega^{2}
r^{2}$ as the potential energy of a body in the tube and in the Schr\"odinger equation

$$i\hbar\frac{\partial \psi}{\partial t} =
-\frac{\hbar^{2}}{2m}\frac{d^{2}\psi}{dr^{2}} + V(r)\psi(r) \eqno(35)$$

However, potential $V(r)$ in the rotating tube is the potential of
the harmonic oscillator with frequency $i\Omega$. All formulas of
harmonic oscillator are here mathematically valid only the problem is that they involve imaginary frequency which cannot be physically
correctly interpreted in classical physics. In other words it is necessary to consider $\psi$ as a wave in the rotating wave guide. It means to consider the Schr\"odinger equation in the rotating tube. It is equivalent to consider the Schr\"odinger equation with the potential of the harmonic oscillators with the imaginary frequenc $\Omega$. The equation for the stationary states is then as it follows:

$$ \frac{d^{2}\psi}{dr^{2}} + \frac {2m}{\hbar^{2}}\left(E + \frac{m\Omega^{2}r^{2}}{2} \right)\psi = 0\eqno(36)$$

The corresponding energies of the "stationary" states are:

$$E_{n} =  -i\hbar\Omega(n + \frac{1}{2})\eqno(37)$$ 

So, we see that the eigenvalue-problem leads to the states which are decaying. The classical limit of the solution is evidently the motion of a charged particle accelerating by potential $V(r) = \frac{1}{2}\Omega^{2}r^{2}$.

\section{Discussion}

We have presented the Lagrange theory of the non-inertial classical
systems, classical particle motion and spin motion in the rotating tube and quantum motion in the rotating tube.
There are  other effects which is possible to consider.
For instance, M\"ossbauer effect in the rotating tube, Pound-Rebka effect in the rotating tube, the \v Cerenkov effect in the
rotating dielectric tube, conductivity and superconductivity in the rotating tube and so on. 

The solved problems were not involved in the framework of the GRG. On the other hand GRG is able to define geometry on the rotating
disk which cannot be composed from the rotating ribbons or
nanoribbons.  Let us discuss geometry of he rotating disk and show
some physical consequences which differs from the physics in the rotating tube.

If we use the the Minkowski element
$$ds^2  =  -c^2{dt'}^2  + {dx'}^2 +  {dy'}^2 + {dz'}^2 \eqno(38)$$
and the nonrelativistic transformation to the rotation system (Matsuo, 2011)

$$d{\bf r'} = d{\bf r} + ({\bf \Omega\times{\bf r}})dt\eqno(39)$$
then we get that space-time element can be expressed in the vectorial form:

$$ds^2  =  g_{\mu\nu}dx^{\mu}dx^{nu }  = 
[-c^2 + ({\bf \Omega\times{\bf r}})^2 ]{dt}^2  + (d{\bf r})^2   +
  2({\bf \Omega\times{\bf r}})dtd{\bf r}.\eqno(40)$$

Thus the metric in the rotating frame can be written by the matrix:

$$g_{\mu\nu} = \left(\begin{array}{cccc}
-1+{\bf u}(x)^{2} &u _{x} & u_{y} & u_{z}\\
u_{x} & 1 & 0 & 0\\
u_{y} & 0 & 1 & 0\\
u_{z} & 0 & 0 & 1\\
\end{array} \right) 
,\eqno(41)$$
where

$${\bf u}(x) = {\Omega(t)}\times{\bf r}/c.\eqno(42)$$

Matsuo et al. (2011) applied the derived metric in order to derive the Dirac equation in the rotating system in order to solve the quantum mechanical problems of the spin-dependent inertial force and spin current in accelerating system. Nevertheless, the knowledge of space-time metric of the rotation system leads to the results which cannot be involved in the physics of the rotated tube because of different formalism, as can be easily seen.  

The transformation between inertial and rotation
system is necessary because it enables
to describe the motion of the particle and spin in the LHC by the
general relativistic methods. The basic idea is the generalization of the so called Lorentz equation for the charged particle in the
electromagnetic field $F^{\mu\nu}$ (Landau et al., 1988):

$$mc \frac{dv^{\mu}}{ds} = \frac{e}{c}F^{\mu\nu}v_{\nu}.\eqno(43)$$

In other words, the normal derivative must be replaced by the covariant one and
we get the general relativistic equation for the motion of a charged
particle in the electromagnetic field and gravity (Landau et al.,
1988):

$$ mc \left(\frac{dv^{\mu}}{ds} +
 \Gamma^{\mu}_{\alpha\beta}v^{\alpha}v^{\beta}\right)
 = \frac{e}{c}F^{\mu\nu}v_{\nu}, \eqno(44)$$
where

$$\Gamma^{\mu}_{\alpha\beta} = \frac{1}{2}g^{\mu\lambda}
\left(\frac{\partial g_{\lambda\alpha}}{\partial x^{\beta}} +
\frac{\partial g_{\lambda\beta}}{\partial x^{\alpha}} -
\frac{\partial g_{\alpha\beta}}{\partial x^{\lambda}}
\right)\eqno(45)$$
are the Christoffel symbols derived in the Riemann geometry theory
(Landau et al., 1988).

In case that we consider motion in the
rotating system, then it is necessary to insert the metrical
tensor $g_{\mu\nu}$, following from the Minkowski element for the
rotation system. The construction of LHC with orbiting protons  must
be in harmony with equation (44) because orbital protons respect the
Coriolis force caused by the rotation of the Earth.

The analogical situation occurs for the motion of the spin. While the original Bargmann-Michel-Telegdi equation for the spin motion is as follows (Berestetzkii et al., 1988)

$$ \frac{da^{\mu}}{ds} = 2\mu F^{\mu\nu}a_{\nu} - 2\mu' v^{\mu}
F^{\alpha\beta}v_{\alpha}a_{\beta}, \eqno(46)$$
where $\mu' = \mu - e/2m$ and $a_{\mu}$ is the axial vector, which
follows also from the classical limit of the Dirac equation with $\bar\psi i\gamma_{5}\gamma_{\mu}\psi\rightarrow a_{\mu}$ (Rafanelli et al., 1964; Pardy, 1973),
the general relativistic generalization of the Bargmann-Michel-Telegdi
equation  can be obtained by the analogical procedure which was
performed with the Lorentz equation. Or,

$$ \left(\frac{da^{\mu}}{ds} +
  \Gamma^{\mu}_{\alpha\beta}v^{\alpha}a^{\beta}\right)
= 2\mu F^{\mu\nu}a_{\nu} - 2\mu' v^{\mu}
F^{\alpha\beta}v_{\alpha}a_{\beta}, \eqno(47)$$
where in case of the rotating system the metrical tensor $g_{\mu\nu}$
must be replaced by the metrical tensor of the rotating system. Then,
the last equation will describe the motion of the spin in the rotating
system.

The motion of the polarized proton in LHC will be described by the last equation because our Earth rotates. During the derivation we wrote $\Gamma^{\mu}_{\alpha\beta}v^{\alpha}a^{\beta}$
and not $\Gamma^{\mu}_{\alpha\beta}v^{\alpha}v^{\beta}$, because every
term must be the axial vector. In other words, the last
equation for the motion of the spin in the rotating system was not
strictly derived but created with regard to the philosophy of author that physics is based on creativity and logic.

On the other hand, the equation (47) must evidently follow from the
Dirac equation in the rotating system, by the same WKB methods which
were  used by  Rafanelli, Schiller and Pardy
( Rafanelli and  Schiller, 1964; Pardy, 1973).
The derived BMT equation in the metric  of the rotation of the Earth
are fundamental for the proper work of LHC because every orbital
proton of  LHC respects the rotation of the Earth and every orbital proton spin respects the Earth rotation too. 

We hope that the named problems are interesting and their
solution will be integral part of the theoretical physics.

\vspace{5mm}

{\bf References}

\vspace{5mm}

\noindent
Berestetzkii, V. B., Lifshitz, E. M. and Pitaevskii, L. P. (1999).
{\it Quantum electrodynamics}, (Butterworth-Heinemann, Oxford).\\[2mm]
Euler, L. 1790).{\it Teoria motus corporum solidorum seu rigidorum}, Editio Nova, Gryphiswaldiae, MDCCXC (1790). \\[2mm]
Frenkel, J. I., (1926). Die Elektrodynamik der rotierenden Elektronen, Zs. Physik {\bf 37}, 243. \\[3mm]
Frenkel, J. I., (1958). {\it Collective scientific works}, II., {\it Scientific articles}, AN SSSR, (in Russian).\\[3mm]
Landau, L. D. and Lifshitz, E. M. (1965). {\it Mechanics}, (Moscow,
Nauka), (in Russian). \\[2mm]
Landau, L. D. and Lifshitz, E. M. (2000). {\it The Classical Theory of Fields}, (Butterworth-Heinemann, Oxford). \\[2mm]
Matsuo, M., Ieda, J., Saitoh, E. and Maekawa, S. (2011). 
Spin-dependent inertial force and spin current in accelerating systems:1106.0366v1 [cond-mat.mes-hall] 2 Jun 2011 \\[2mm]
Pardy, M. (1973). Classical motion of spin 1/2 particles with zero
anomalous magnetic moment, {Acta Physica Slovaca} {\bf 23},
No. 1, 5. \\[2mm]
Pardy, M. (2007). Bound motion of a bodies and particles in the rotating Systems: Revised and modifed version of astro-ph/0601365, International Journal of Theoretical Physics, Vol. 46, No. 4, April 2007 \\[2mm]
Pardy, M. (2009). The bremsstrahlung equation for the spin motion  in  electromagnetic field: International Journal of Theoretical Physics, Volume 48,Number 11 / November, 2009, pp. 3241-3248, DOI10.1007/s10773-009-0127-6.\\[2mm]
Rafanelli, K, and Schiller, R. (1964). Classical motion of
spin-1/2 particles, {\it Phys. Rev.} {\bf 135}, No. 1 B, B279.
\\[3mm]
Ternov, I. M., (1980). On the contemporary interpretation of the classical theory of the J. I. Frenkel spin, Uspekhi fizicheskih nauk, {}{\bf 132}, 345. (in Russian). \\[3mm]    
Thomas, L. H., (1926). The motion of spinning electron, Nature, {\bf 117}, 514. \\[3mm]
Tomonaga, S.-I., (1997). {\it The story of spin}, (The university of Chicago press, Ltd., London).\\[3mm] 

\end{document}